\documentclass[12pt]{iopart}

\usepackage{bm,graphicx}
\newcommand{\figwidth}{3.in}
\newcommand{\simleq}{\hspace{0.3em}\raisebox{0.4ex}{$<$}\hspace{-0.75em}\raisebox{-.7ex}{$\sim$}\hspace{0.3em}}

\begin{document}

\title[Time evolution of autocorrelation function in dynamical replica theory]
{Time evolution of autocorrelation function in dynamical replica
theory}

\author{A Sakata}

\address{
Department of Computational Intelligence and Systems Science,
Tokyo Institute of Technology,
Midori-ku, Yokohama 226-8502, Japan}
\ead{ayaka@sp.dis.titech.ac.jp}
\begin{abstract}
Asynchronous dynamics given by the master equation in the
Sherrington--Kirkpatrick (SK) spin-glass model is studied based on
dynamical replica theory (DRT) with an extension to take into
account the autocorrelation function.
The dynamical behaviour of the system is approximately described by
dynamical equations of the macroscopic quantities: magnetization,
energy contributed by randomness, and the autocorrelation function.
The dynamical equations under the replica symmetry assumption are
derived by introducing the subshell equipartitioning assumption and
exploiting the replica method. The obtained dynamical equations are
compared with Monte Carlo (MC) simulations, and it is demonstrated
that the proposed formula describes well the time evolution of the
autocorrelation function in some parameter regions.
The study offers a reasonable description of the autocorrelation
function in the SK spin-glass system.
\end{abstract}


\section{Introduction}

Random systems that possess frustration have many metastable states
in their free energy landscape, which are expected to contribute to
the system behaviour. Spin-glass systems are representative of such
systems and their equilibrium properties have been extensively
studied in a mean-field model, the Sherrington--Kirkpatrick (SK)
model. Mean-field theory based on the replica method reveals the
existence of a transition associated with replica symmetry breaking
(RSB) that is interpreted with the aid of Thouless--Anderson--Palmer
(TAP) theory as the exponential appearance of metastable states in
the free energy landscape \cite{beyond,Nishimori}. A quantitative
understanding of the metastable states is attempted by counting the
number of solutions of the TAP equation or belief propagation
equation \cite{Bray-Moore,Monasson,Aspelmeier-Bray-Moore}.

Below a critical temperature, spin glasses in experiments show slow relaxation
dynamics and an aging effect in which the dynamical properties
depend on the history even after a long time period \cite{aging}.
The dynamical behaviour and equilibrium properties are considered to
be related, but the connection is not fully understood.
It has been suggested that dynamical behaviour of
the SK model is qualitatively similar to spin glasses in experiments even
though the realistic spin glasses have finite dimensions
\cite{Cugliandolo-Kurchan-Ritort}. Therefore, clarifying the
relaxation dynamics in the SK model provides insight into the
connection between the equilibrium and dynamical properties, as well
as into the dynamics of random systems.
For this purpose, an analytical description of the relaxation dynamics
in the SK model that can be compared with our knowledge of the equilibrium states would be of significant use.

Dynamical replica theory (DRT) provides a statistical-mechanical
description of the dynamics in random systems
\cite{Coolen-Sherrington}. In DRT, the time evolution of macroscopic
quantities is derived according to the replica method. A
characteristic feature of DRT is the introduction of an assumption
called subshell equipartitioning in which the microscopic states in
the subshell that are characterized by a certain macroscopic quantity
value appear with the same probability. By introducing this
assumption, one can obtain closed dynamical equations that describe
the time evolution of the macroscopic quantities. The derivation of
the dynamical equations and their DRT solutions are tractable even
if the relaxation dynamics over long time are considered, although an exact
description of the dynamics is sacrificed. DRT offers a contrasting
perspective compared to the generating functional method
\cite{deDominicis,Hatchett} in which the dynamical equations are
exact but difficult to solve over long time steps.

The physical implications of subshell equipartitioning and its
influence on the accuracy of the theoretical prediction have been
studied \cite{Tanaka-Osawa,Ozeki-Nishimori}. The assumption removes
the microscopic time correlation, and the system only depends on the
past value of the macroscopic quantities that characterize the
subshell. It is well known that the time correlation is significant
in describing the dynamics of random systems. In fact, it has been
demonstrated that analytical descriptions of the dynamics are
improved by taking into account the time correlation in the Hopfield
model \cite{Amari-Maginu,Okada} and in the interference canceller of
code-division-multiple-access (CDMA) multiuser detection
\cite{Tanaka-Okada}. However, DRT in its present formalism cannot
deal with the time correlation, and hence to apply DRT to random
systems and discuss the time correlation, an extension should be
introduced.

A central quantity in the study of the dynamics of spin glasses is
the spin autocorrelation function that describes the correlation between
microscopic states at two time steps. It is experimentally and numerically
observed with the conjugate response function as a quantity that
indicates the slow relaxation dynamics \cite{aging} and its
analytical description has been studied in particular Langevin
dynamics of soft spins
\cite{Sompolinsky-Zippelius,Cugliandolo-Kurchan}. In this paper, I
study the sequential dynamics of the SK model with DRT while
including the time evolution of the autocorrelation function. DRT
has been studied mainly for two cases where the subshell is
characterized by a magnetization and energy set contributed by
randomness \cite{Coolen-Sherrington} or a conditional local field
distribution \cite{Laughton}. I start with the simplest DRT
\cite{Coolen-Sherrington} and introduce the autocorrelation function
into the formulation. The evolution equation of the autocorrelation
function is derived based on the replica method and its dependence
on the phase is discussed.

This paper is organized as follows. In section \ref{Model}, I
explain the model settings and time evolution of the joint
probability of microscopic states at two time steps, which is
described by Glauber dynamics \cite{Glauber}. In section
\ref{Time evolution}, macroscopic quantities, magnetization, energy
contributed by randomness, and the autocorrelation function, are
introduced to characterize the microscopic state space. The closed
formula of the evolutional dynamics of the quantities is obtained
following the procedures of DRT, and the expressions under the
replica symmetry (RS) assumption are derived. In section
\ref{results}, the obtained dynamical equations are numerically
solved and the results are compared with MC simulations.
Finally, section \ref{Summary} is devoted to conclusions and the
outlook for further developments.

\section{Model setup}
\label{Model}

The system being studied here is a mean-field spin-glass model, the
SK model, which consists of $N$ Ising spins
$\bm{S}=\{S_i\}~(i=1,\cdots,N)$. The Hamiltonian is given by
\begin{eqnarray}
H(\bm{S}|\bm{J})=-\sum_{i<j}J_{ij}S_iS_j-h\sum_i S_i,
\label{eq:Hamiltonian}
\end{eqnarray}
where $J_{ij}$ is the interaction between spins $S_i$ and $S_j$, and
$h$ is the external field. The interactions $\bm{J}=\{J_{ij}\}$ are
symmetric ($J_{ij}=J_{ji}$) and assumed to be independently and
identically distributed according to the Gauss distribution with a
mean of $J_0\slash N$ and a variance of $J^2\slash N$;
\begin{eqnarray}
P_J(J_{ij})=\sqrt{\frac{N}{2\pi
 J^2}}\exp\Big[-\frac{N}{2J^2}\Big(J_{ij}-\frac{J_0}{N}\Big)^2\Big].~~~
 (i\neq j)
\label{eq:P_J}
\end{eqnarray}

Here, I consider the conditional probability of a microscopic spin
configuration $\bm{\sigma}$ at $t_w+t$ under a given spin
configuration $\bm{s}$ at time $t_w$ and interaction $\bm{J}$ as
$p(\bm{\sigma};t_w+t|\bm{s};t_w,\bm{J})$. The time $t_w$ is called
the waiting time. The two configurations $\bm{\sigma}$ and $\bm{s}$
are related to each other through $\bm{\sigma}=\bm{s}\circ\bm{d}$,
where $\bm{d}=\{d_i\}~(i=1,\cdots,N)$ is a vector consisting of
Ising variables, and $\circ$ denotes the Hadamard product, which is
a product with respect to each component; the $i$-th component of
$\bm{s}\circ\bm{d}$ is $s_id_i$. When $d_i=1$, the $i$-th spin
configuration at time $t_w$ and that at time $t_w+t$ are the same;
otherwise they differ. The time evolution of the microscopic
conditional probability in the direction of $t$ is given by Glauber
dynamics, which is an asynchronous update of spin configurations
described by the master equation,
\begin{eqnarray}
\nonumber
\frac{d}{dt}p(\bm{s}\circ\bm{d};t_w+t|\bm{s};t_w,\bm{J})&=\sum_{k=1}^N\{p(\bm{s}\circ F_k\bm{d};t_w+t|\bm{s};t_w,\bm{J})w_k(\bm{s}\circ F_k\bm{d}|\bm{J})\\
&\hspace{1.0cm}-p(\bm{s}\circ \bm{d};t_w+t|\bm{s};t_w,\bm{J})w_k(\bm{s}\circ \bm{d}|\bm{J})\},
\label{eq:Glauber}
\end{eqnarray}
where $F_k$ is an operator that flips the configuration of the
$k$-th Ising variable as
$F_k\bm{s}=\{s_1,\cdots,-s_k,\cdots,s_N\}$. The transition
probability under a given $\bm{J}$, $w_k(\bm{s}|\bm{J})$, is given
by
\begin{eqnarray}
w_k(\bm{s}|\bm{J})=\frac{1}{2}(1-s_k\tanh(\beta h_k(\bm{s}|\bm{J}))),
\label{eq:transition}
\end{eqnarray}
where $h_k(\bm{s}|\bm{J})=\sum_{i\neq k}J_{ik}s_i+h$ is the local
field of the $k$-th spin. The time evolution of the joint
probability $p(\bm{\sigma};t_w+\tau,\bm{s};t_w|\bm{J})$ is obtained by
multiplying both sides of \eref{eq:Glauber} by the probability
distribution of a microscopic state at time $t_w$, $p(\bm{s};t_w|\bm{J})$.
The fixed point of \eref{eq:Glauber} that is obtained at
$t\to\infty$ corresponds to the equilibrium distribution
\begin{eqnarray}
p(\bm{\sigma};t_w+t,\bm{s};t_w|\bm{J})\to \frac{1}{Z}p(\bm{s};t_w|\bm{J})\exp(-\beta H(\bm{\sigma}|\bm{J})),
\end{eqnarray}
where $Z$ is the normalization constant.

\section{Time evolution of macroscopic quantities}
\label{Time evolution}

The macroscopic states are defined by the following macroscopic quantities,
\begin{eqnarray}
m_0(\bm{s})&=\frac{1}{N}\sum_{i=1}^Ns_i,~~r_0(\bm{s})=\frac{1}{N}\sum_{i<j}\Big(J_{ij}-\frac{J_0}{N}\Big)s_is_j\\
m(\bm{\sigma})&=\frac{1}{N}\sum_{i=1}^N\sigma_i,~~r(\bm{\sigma})=\frac{1}{N}\sum_{i<j}\Big(J_{ij}-\frac{J_0}{N}\Big)\sigma_i\sigma_j\\
c(\bm{d})&=\frac{1}{N}\sum_{i=1}^N\sigma_is_i=\frac{1}{N}\sum_{i=1}^Nd_i,
\end{eqnarray}
where $m_0$ and $r_0$ are the magnetization and the absolute value
of the energy contributed by randomness (referred to as randomness
energy, hereafter) at time $t_w$, respectively, and $m$ and $r$ are
the corresponding values at time $t_w+t$. The quantity $c$ is the
autocorrelation function between time $t_w+t$ and $t_w$, and it is a
function of $\bm{d}$. The probability distribution of the
macroscopic quantities is given by
\begin{eqnarray}
\nonumber
P_{t_w+t,t_w}(\Omega|\bm{J})&=\sum_{\bm{\sigma},\bm{s}}p(\bm{\sigma};t_w+t,\bm{s};t_w|\bm{J})\delta(m_0-m(\bm{s}))\delta(r_0-r(\bm{s}))\\
&\hspace{1.0cm}\times\delta(m-m(\bm{\sigma}))\delta(r-r(\bm{\sigma}))\delta(c-c(\bm{d})),
\label{eq:prob-dist}
\end{eqnarray}
where $\Omega\equiv\{m_0,r_0,m,r,c\}$. There are three Ising
variables $\bm{\sigma}$, $\bm{s}$, and $\bm{d}$, but the
relationship $\bm{\sigma}=\bm{s}\circ\bm{d}$ reduces the number of
the independent variables to two. The double summation over the
microscopic states of $\bm{\sigma}$ and $\bm{s}$ should be taken in
mind of the relationship $\bm{\sigma}=\bm{s}\circ\bm{d}$. The time
evolution of $P_{t_w+t,t_w}(\Omega|\bm{J})$ in the direction of $t$
is derived by substituting (\ref{eq:Glauber}) and
(\ref{eq:transition}) into (\ref{eq:prob-dist}) and employing the
Taylor expansion with respect to $1\slash N$:
\begin{eqnarray}
\nonumber
\frac{d}{dt}&P_{t_w+t,t_w}(\Omega|\bm{J})\\
\nonumber
=&-\frac{\partial}{\partial m}P_{t_w+t,t_w}(\Omega|\bm{J})\Big(\frac{1}{N}\sum_{k=1}^N\langle\langle\tanh\beta(J_0m+Jz_k(\bm{\sigma}|\bm{J})+h)\rangle\rangle_\Omega-m\Big)\\
\nonumber
&-\frac{\partial}{\partial
r}P_{t_w+t,t_w}(\Omega|\bm{J})\Big(\frac{1}{N}\sum_{k=1}^N\langle\langle z_k\tanh\beta(J_0m+Jz_k(\bm{\sigma}|\bm{J})+h)\rangle\rangle_\Omega-2r\Big)\\
&-\frac{\partial}{\partial
 c}P_{t_w+t,t_w}(\Omega|\bm{J})\Big(\frac{1}{N}\sum_{k=1}^N\langle\langle s_k\tanh\beta(J_0m+Jz_k(\bm{\sigma}|\bm{J})+h)\rangle\rangle_\Omega-c\Big),
\label{eq:dP/dt}
\end{eqnarray}
where 
$z_k(\bm{\sigma}|\bm{J})=\sum_{i\neq
k}(J_{ki}-J_0\slash N)\sigma_i$ is the local field contributed by
randomness of the $k$-th spin. The notation
$\langle\langle\cdots\rangle\rangle_\Omega$ represents the average
within the subshell specified by the set of macroscopic quantities
$\Omega$ as
\begin{eqnarray}
\langle\langle\cdots\rangle\rangle_\Omega=\frac{\sum_{\bm{\sigma},\bm{s}}\cdots
 p(\bm{\sigma};t_w+t,\bm{s};t_w|\bm{J}){\cal W}(\bm{\sigma},\bm{s}|\Omega)}{\sum_{\bm{\sigma},\bm{s}}p(\bm{\sigma};t_w+t,\bm{s};t_w|\bm{J}){\cal W}(\bm{\sigma},\bm{s}|\Omega)},
\end{eqnarray}
where ${\cal
W}(\bm{\sigma},\bm{s}|\Omega)\equiv\delta(m_0-m_0(\bm{s}))\delta(r-r_0(\bm{s}))\delta(m-m(\bm{\sigma}))\delta(r-r(\bm{\sigma}))\delta(c-c(\bm{d}))$.
By regarding \eref{eq:dP/dt} as the total differential form with
respect to $m,~r$, and $c$, their dynamical equations are given by
\begin{eqnarray}
\frac{d}{dt}m&=\int dz~\sum_{S=\pm 1}D(z,S|\Omega,\bm{J})\tanh(\beta(J_0m+Jz+h))-m\label{eq:dm/dt}\\
\frac{d}{dt}r&=\int dz~\sum_{S=\pm 1}D(z,S|\Omega,\bm{J})\tanh(\beta(J_0m+Jz+h))-2r\label{eq:dr/dt}\\
\frac{d}{dt}c&=\int dz~\sum_{S=\pm 1}D(z,S|\Omega,\bm{J})S\tanh(\beta(J_0m+Jz+h))-c,\label{eq:dc/dt}
\end{eqnarray}
where $S$ takes the value $\pm 1$, and $z$ is the effective noise
caused by quenched randomness. The effective noise is distributed
according to $D(z,S|\Omega,\bm{J})$, which is given by
\begin{eqnarray}
D(z,S|\Omega,\bm{J})=\lim_{N\to\infty}\frac{\sum_{\bm{\sigma},\bm{s}}\frac{1}{N}\sum_k\delta(z-z_k(\bm{\sigma}|\bm{J}))\delta_{s_k,S}P_{t_w+t,t_w}(\Omega|\bm{J})}{\sum_{\bm{\sigma},\bm{s}}P_{t_w+t,t_w}(\Omega|\bm{J})}
\end{eqnarray}
where $\delta_{s_k,S}$ is the Kronecker delta.

\subsection{Closed formula of the dynamical equations}

Following the procedures of DRT, a closed formula of the dynamical
equations of the macroscopic variables is obtained by assuming the
realization of two properties: self-averaging and subshell
equipartitioning. The self-averaging property guarantees that the
noise distribution, $D(z,S|\Omega,\bm{J})$, which depends on a
realization of $\bm{J}$, converges to a typical distribution
$D(z,S|\Omega)\equiv[D(z,S|\Omega,\bm{J})]_J$ in the limit
$N\to\infty$, where $[\cdots]_J$ is the average over the randomness
$\bm{J}$ according to the probability distribution \eref{eq:P_J}.
Subshell equipartitioning ensures that the microscopic states within
a subshell specified by the macroscopic quantities
$\Omega=\{m_0,r_0,m,r,c\}$ appear with the same probability at each
time step. As a consequence of these assumptions, the noise
distribution $D(z,S|\Omega,\bm{J})$ is replaced with $D(z,S|\Omega)$
as given by
\begin{eqnarray}
D(z,S|\Omega)=\lim_{N\to\infty}\Big[\frac{\sum_{\bm{\sigma},\bm{s}}\frac{1}{N}\sum_k
\delta(z-z_k(\bm{\sigma}|\bm{J}))\delta_{s_k,S}{\cal W}(\bm{\sigma},\bm{s}|\Omega)}{\sum_{\bm{\sigma},\bm{s}}{\cal W}(\bm{\sigma},\bm{s}|\Omega)}\Big]_J.
\label{eq:D_assumption}
\end{eqnarray}

At equilibrium, the probability distribution of the microscopic
states is given by the Boltzmann distribution with a Hamiltonian
\eref{eq:Hamiltonian}
that can be expressed in terms of $m$ and $r$, and
hence the subshell equipartitioning property with $m$ and $r$ is
realized. However, the property is not expected to hold in cases far
from equilibrium, and to apply the subshell equipartitioning
assumption to non-equilibrium states is crucial for the theoretical
accuracy.

The average over the quenched randomness in \eref{eq:D_assumption}
is calculated by introducing the replica expression as
\begin{eqnarray}
\nonumber
D(z,S|\Omega)&=\lim_{n\to
 0}\lim_{N\to\infty}\sum_{\bm{\sigma}^1,\cdots,\bm{\sigma}^n}\sum_{\bm{s}^1,\cdots,
 \bm{s}^n}\Big[\frac{1}{N}\sum_{k=1}^N\delta(z-z_k(\bm{\sigma}^1))
 \delta_{s_k^1,S}\\
&\hspace{5.0cm}\times\prod_{a=1}^n{\cal W}(\bm{\sigma}^a,\bm{s}^a|\Omega)\Big]_J.
\label{eq:D_replica}
\end{eqnarray}
The right-hand side of \eref{eq:D_replica} is calculated for a
positive integer $n$ and then analytically continued to a
non-integer $n$ by taking the limit to 0. After the calculations,
the noise distribution can be expressed as
\begin{eqnarray}
\nonumber
D(z,S|\Omega)
&=\lim_{n\to0}\lim_{N\to\infty}\frac{1}{\sqrt{2\pi}}e^{N\Phi-{\rm P}}
\langle\delta_{\sigma_0^1,S}\exp\{-\frac{1}{2}(z-\overline{z})^2\}\rangle_{\Xi}
\end{eqnarray}
where ${\rm P}=\frac{1}{4}\sum_{ab}(\rho^a\rho^b+\rho^a_0\rho^b+\rho^b_0\rho^a+\rho^a_0\rho^b_0)$,
$\overline{z}=\rho^1\sigma^1+\rho_0^as^ac+\sum_{a=2}^n(\rho^a
 \sigma^a q^{1a}+\rho_0^a s^a q^{1a}_{t})$,
\begin{eqnarray}
\nonumber
\Phi&=\mathop{\rm
extr}_{\hat\Omega,{\cal Q}}\Big[-\sum_{a=1}^n(m\mu^a+m_0\mu^a_0+r\rho^a+r_0
\rho^a+c\gamma^a)+\frac{1}{4}\sum_a({\rho^a}^2+2c^2\rho^a\rho^a_0+{\rho^a_0}^2)\\
&-\frac{1}{2}\sum_{a<b}(\rho^\alpha {q^{ab}}^2\rho^b+\rho_0^a{q^{ab}_0}^2\rho_0^b+\rho^a_0{q^{ab}_t}^2\rho^b+\rho^b_0{q^{ba}_t}^2\rho^a)
+\log\sum_{\bm{\sigma},\bm{s}}e^{\Xi}\Big],\label{eq:Phi}\\
\nonumber
\Xi&=
 \sum_{a}(\mu^a\sigma^a+\mu^a_0s^a+\gamma^a
\sigma^a s^a)\\
&\hspace{1.0cm}+\sum_{a<b}(q^{ab}\rho^{a}\rho^{b}
\sigma^a\sigma^b+q^{ab}_0\rho^{a}_0\rho^{b}_0s^a s^b+q^{ab}_{t}\rho^{a}_0\rho^{b}s^a\sigma^b+q^{ba}_{t}\rho^{b}_0\rho^{a}s^b\sigma^a),
\label{eq:Xi}
\end{eqnarray}
and $\langle
A\rangle_{\Xi}=\sum_{\{\sigma^a\},\{s^a\}}Ae^{\Xi}\slash\sum_{\{\sigma^a\},\{s^a\}}e^{\Xi}$.
The variables
$\hat\Omega\equiv\{\{\mu^a_0\},\{\rho_0^a\},\{\mu^a\},\{\rho^a\},\{\gamma^a\}\}$
and ${\cal
Q}\equiv\{\{q^{ab}_0\},\{q^{ab}\},\{q^{ab}_t\}\}~(a,b=1,\cdots,n,a\neq
b)$ are the conjugates of $\Omega=\{m_0,r_0,m,r, c\}$ and the
overlap parameters, respectively. The notation $\mathop{\rm
extr}_{\hat\Omega,{\cal Q}}{\cal F}(\hat\Omega,{\cal Q})$ denotes
the extremization of a function ${\cal F}(\hat\Omega,{\cal Q})$ with
respect to $\hat\Omega$ and ${\cal Q}$. At the extremum of
\eref{eq:Phi}, the variables are related to the physical quantities
through
\begin{eqnarray}
m&=\frac{1}{n}\sum_a\frac{\sum_{\bm{\sigma},\bm{s}}\sigma^a e^\Xi}{\sum_{\bm{\sigma},\bm{s}}e^\Xi},~~m_0=\frac{1}{n}\sum_a\frac{\sum_{\bm{\sigma},\bm{s}}s^a e^\Xi}{\sum_{\bm{\sigma},\bm{s}}e^\Xi},\label{eq:m}\\
r&= \frac{1}{2n}\sum_{ab}({q^{ab}}^2\rho^b+{q^{ab}_{t}}^2\rho^b_0),~~~r_0 = \frac{1}{2n}\sum_{ab}({q^{ab}_0}^2\rho^b_0+{q^{ab}_{t}}^2\rho^b)\label{eq:r}\\
c&=\frac{1}{n}\sum_a\frac{\sum_{\bm{\sigma},\bm{s}}\sigma^as^ae^\Xi}{\sum_{\bm{\sigma},\bm{s}}e^\Xi},\label{eq:c}
\end{eqnarray}
and
\begin{eqnarray}
q^{ab}=\frac{\sum_{\bm{\sigma},\bm{s}}\sigma^a\sigma^be^\Xi}{\sum_{\bm{\sigma},\bm{s}}e^\Xi},~~~q^{ab}_0=\frac{\sum_{\bm{\sigma},\bm{s}}s^as^be^\Xi}{\sum_{\bm{\sigma},\bm{s}}e^\Xi},~~~q^{ab}_{t}=\frac{\sum_{\bm{\sigma},\bm{s}}\sigma^as^be^\Xi}{\sum_{\bm{\sigma},\bm{s}}e^\Xi}.\label{eq:q}
\end{eqnarray}

\subsection{Replica symmetry}

I introduce the RS assumption here for the conjugate and overlap
parameters as
\begin{eqnarray}
\mu^a = \mu,~~~~\mu^a_0 =\mu_0,~~~~\gamma^a=\gamma,\\
q^{ab}=q,~~~~q^{ab}_0=q_0,~~~~q^{ab}_t=q_t,\\
\rho^a=\rho,~~~~\rho^a_0=\rho_0.
\end{eqnarray}
With these assumptions, (\ref{eq:m})--(\ref{eq:q}) are transformed
to
\begin{eqnarray}
m &= \int DuDv\frac{\tanh X+\tanh X_0\tanh Y}{1+\tanh X_0\tanh X\tanh Y}\label{eq:m_RS}\\
m_0 &=\int DuDv\frac{\tanh X_0+\tanh X\tanh Y}{1+\tanh X_0\tanh X\tanh Y}\label{eq:m0_RS}\\
c &=\int DuDv\frac{\tanh Y+\tanh X_0\tanh X}{1+\tanh X_0\tanh X\tanh Y},\label{eq:c_RS}
\end{eqnarray}
\begin{eqnarray}
\rho&=\frac{2\{r_0(c^2-q_t^2)-r(1-q_0^2)\}}{(c^2-q_t^2)^2-(1-q_0^2)(1-q^2)},\\
\rho_0&=\frac{2\{r(c^2-q_t^2)-r_0(1-q^2)\}}{(c^2-q_t^2)^2-(1-q_0^2)(1-q^2)},
\end{eqnarray}
and
\begin{eqnarray}
q &=\int DuDv\Big(\frac{\tanh X+\tanh X_0\tanh Y}{1+\tanh X_0\tanh X\tanh Y}\Big)^2\label{eq:q_RS}\\
q_0&=\int DuDv\Big(\frac{\tanh X_0+\tanh X\tanh Y}{1+\tanh X_0\tanh X\tanh Y}\Big)^2\label{eq:q0_RS}\\
q_t&=\int DuDv\frac{\tanh X+\tanh X_0\tanh Y}{1+\tanh X_0\tanh X\tanh Y}\cdot\frac{\tanh X_0+\tanh X\tanh Y}{1+\tanh X_0\tanh X\tanh Y},\label{eq:qt_RS}
\end{eqnarray}
where
\begin{eqnarray}
X&=\mu+\rho u\sqrt{q}\\
X^0&=\mu_0+\rho_0u\sqrt{\frac{q_t^2}{q}}+\rho_0 v\sqrt{q_0-\frac{q_t^2}{q}}\\
Y&=\gamma-q_t\rho\rho_0.
\end{eqnarray}
The noise distribution under the RS assumption is given by a
function of the RS quantities as
\begin{eqnarray}
\nonumber
D&(z,S|\Omega)=\int \frac{DuDv}{8\sqrt{2\pi}}\\
\nonumber
&\times\Big\{(1+S)\frac{(1+\tanh
 X_1)(1+\tanh X_1^0)(1+\tanh Y)}{1+\tanh
 X_1\tanh X_1^0\tanh Y}e^{-\frac{1}{2}(z-\overline{z}_1)^2}\\
\nonumber
&+(1-S)\frac{(1+\tanh
 X_2)(1-\tanh X_2^0)(1-\tanh Y)}{1+\tanh
 X_2\tanh X_2^0\tanh Y}e^{-\frac{1}{2}(z-\overline{z}_2)^2}\\
\nonumber
&+(1+S)\frac{(1-\tanh
 X_3)(1+\tanh X_3^0)(1-\tanh Y)}{1+\tanh
 X_3\tanh X_3^0\tanh Y}e^{-\frac{1}{2}(z-\overline{z}_3)^2}\\
&+(1-S)\frac{(1-\tanh
 X_4)(1-\tanh X_4^0)(1+\tanh Y)}{1+\tanh
 X_4\tanh X_4^0\tanh Y}e^{-\frac{1}{2}(z-\overline{z}_4)^2}\Big\},
\end{eqnarray}
which is a mixture of four Gaussian distributions. The means of the
four Gaussian distributions are given by
$\overline{z}_1=\rho(1-q)+\rho_0(c-q_t)$,
$\overline{z}_2=\rho(1-q)-\rho_0(c-q_t)$,
$\overline{z}_3=-\rho(1-q)+\rho_0(c-q_t)$, and
$\overline{z}_4=-\rho(1-q)-\rho_0(c-q_t)$, and the functions $X_i$
and $X_i^0$ for $i=1,\cdots,4$ are given by
\begin{eqnarray}
X_i&=\mu+\rho\sqrt{q}\{\sqrt{1-q}u+\sqrt{q}(z-\overline{z}_i)\},\\
X_i^0&=\mu_0+\rho_0\left\{\frac{q_t}{\sqrt{q}}\left(\sqrt{1-q}u+\sqrt{q}(z-\overline{z}_i)\right)+\sqrt{q_0-\frac{q_t^2}{q}v}\right\}.
\end{eqnarray}

\subsection{Fixed point of the macroscopic equations}
\label{sec:fixed-point}

The dynamical equations reach the fixed point, for which $dm\slash
dt=0$, $dr\slash dt = 0$, and $dc\slash dt = 0$, after a long time
evolution of $t\to\infty$. At the fixed point denoted by $m^*$,
$r^*$, and $c^*$, the following relationships hold:
\begin{eqnarray}
m^*&=\int Dz \lim_{n\to 0}
\langle\tanh\beta(J_0m+J(z+\overline{z})+h)\rangle_{\Xi}\label{eq:m_fix}\\
r^*&=\frac{1}{2}\int Dz\lim_{n\to 0}\left<(z+\overline{z})\tanh\beta(J_0m+J(z+\overline{z})+h)\right>_\Xi \label{eq:r_fix}\\
c^*&=\int Dz \lim_{n\to 0}\langle s^1\tanh\beta(J_0m+J(z+\overline{z})+h)\rangle_{\Xi},\label{eq:c_fix}
\end{eqnarray}
where $\int Dz=(2\pi)^{-1\slash 2}\int_{-\infty}^{\infty}
dz\exp(-z^2\slash 2)$. Equations (\ref{eq:m_fix})--(\ref{eq:c_fix})
are obtained by substituting $D(z,S|\Omega)$ into
(\ref{eq:dm/dt})--(\ref{eq:dc/dt}). When the relationships
$\mu=\beta(J_0m^*+h),~\gamma=\rho\rho_0c^*,~\rho=\beta J$, and
$c^*=q_t$ are satisfied under the RS assumption, the RS fixed point
equations (\ref{eq:m_fix})--(\ref{eq:c_fix}) are transformed as
\begin{eqnarray}
m^*&=\int Du\tanh\beta\left(J_0m^*+h+J\sqrt{q}u\right)\label{eq:m_fix-2}\\
r^*&=\frac{\beta J}{2}\left(1-q^2\right)\label{eq:r_fix-2}\\
\nonumber
c^*&=\int DuDv\tanh\beta\left(J_0m^*+h+J\sqrt{q}u\right)\\
&\hspace{1.0cm}\times\tanh\left(\mu_0+\rho_0(\sqrt{\frac{q_t^2}{q}}u+\sqrt{q_0-\frac{q_t^2}{q}}v)\right).\label{eq:c_fix-2}
\end{eqnarray}
In this case, the overlaps are given by
\begin{eqnarray}
q_0&=\int Du\tanh^2\left(\mu_0+\rho_0\sqrt{q_0}u\right)\\
q&=\int Du\tanh^2\beta\left(J_0m^*+h+J\sqrt{q}u\right),\label{eq:q_fix}
\end{eqnarray}
and the equations to determine the conjugate $\mu_0$ is given by
\begin{eqnarray}
m_0&=\int Du\tanh\left(\mu_0+\rho_0\sqrt{q_0}u\right).
\end{eqnarray}
The conjugates of the energy $r$ and $r_0$ are given by
$\rho=2r^*\slash(1-q^2)$ and $\rho_0=2r_0\slash(1-q_0^2)$,
respectively, and the former relationship indicates the consistency
of $\rho=\beta J$. Equations \eref{eq:m_fix-2}, \eref{eq:r_fix-2},
and \eref{eq:q_fix} correspond to the magnetization, randomness
energy, and overlap in the SK model under the RS assumption,
respectively, and hence the RS solution of the SK model is recovered
as a solution of the fixed point equations, as with the original
DRT.

\section{Results}
\label{results}

\begin{figure}
\begin{center}
\includegraphics[width=\figwidth]{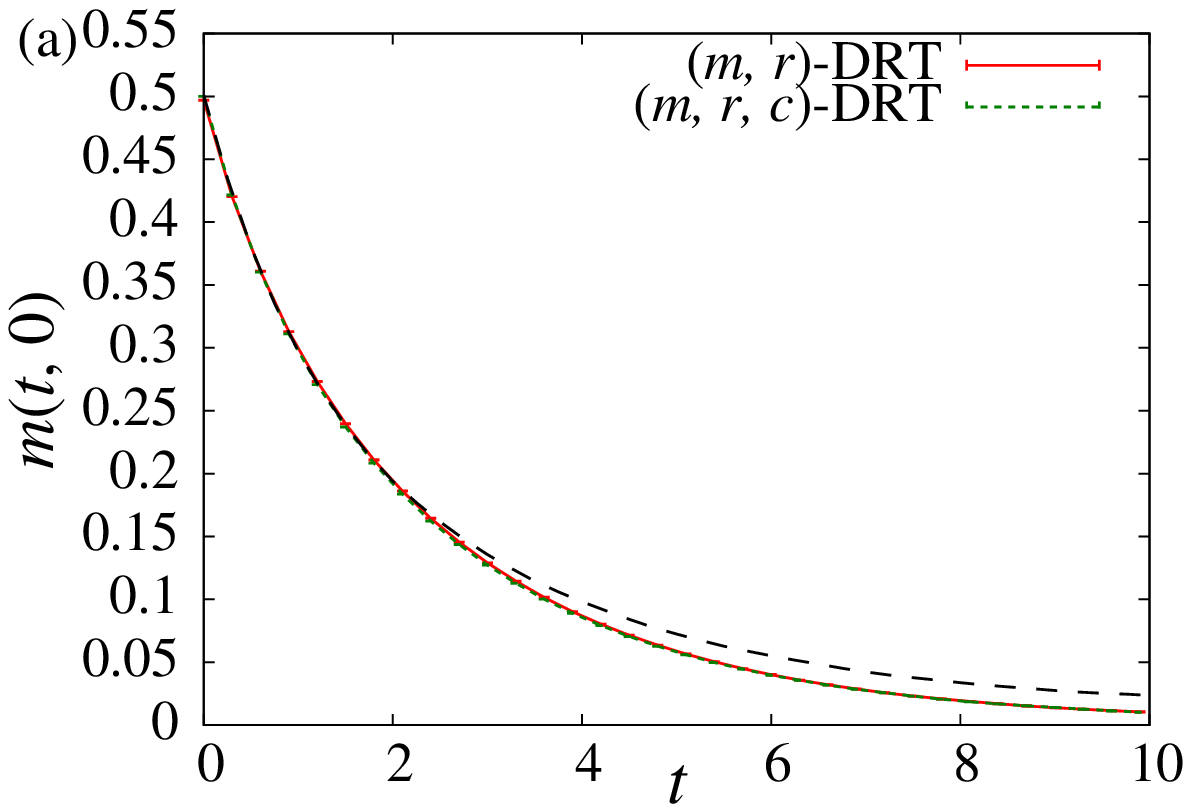}
\includegraphics[width=\figwidth]{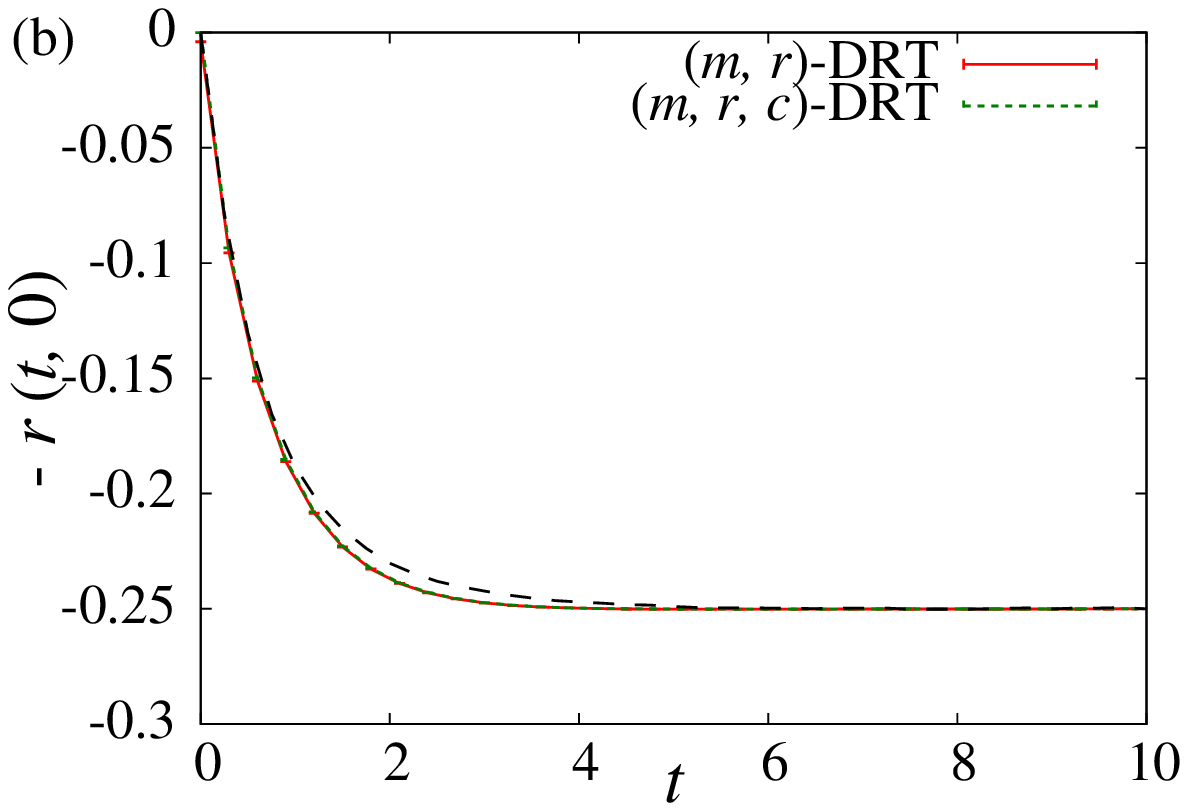}
\end{center}
\caption{(colour online) Time evolution of (a) $m$ and (b) $-r$ at
$T\slash J=2.0$ and $J_0\slash J=1.0$, whose equilibrium state
corresponds to a paramagnetic phase. The solid and dotted lines
represent the dynamics theoretically predicted by the $(m,r)$--DRT
and $(m,r,c)$--DRT, respectively. The dashed line represents the
results of the MC simulation at $N=4096$. } \label{fig:para-mr}
\begin{center}
\includegraphics[width=\figwidth]{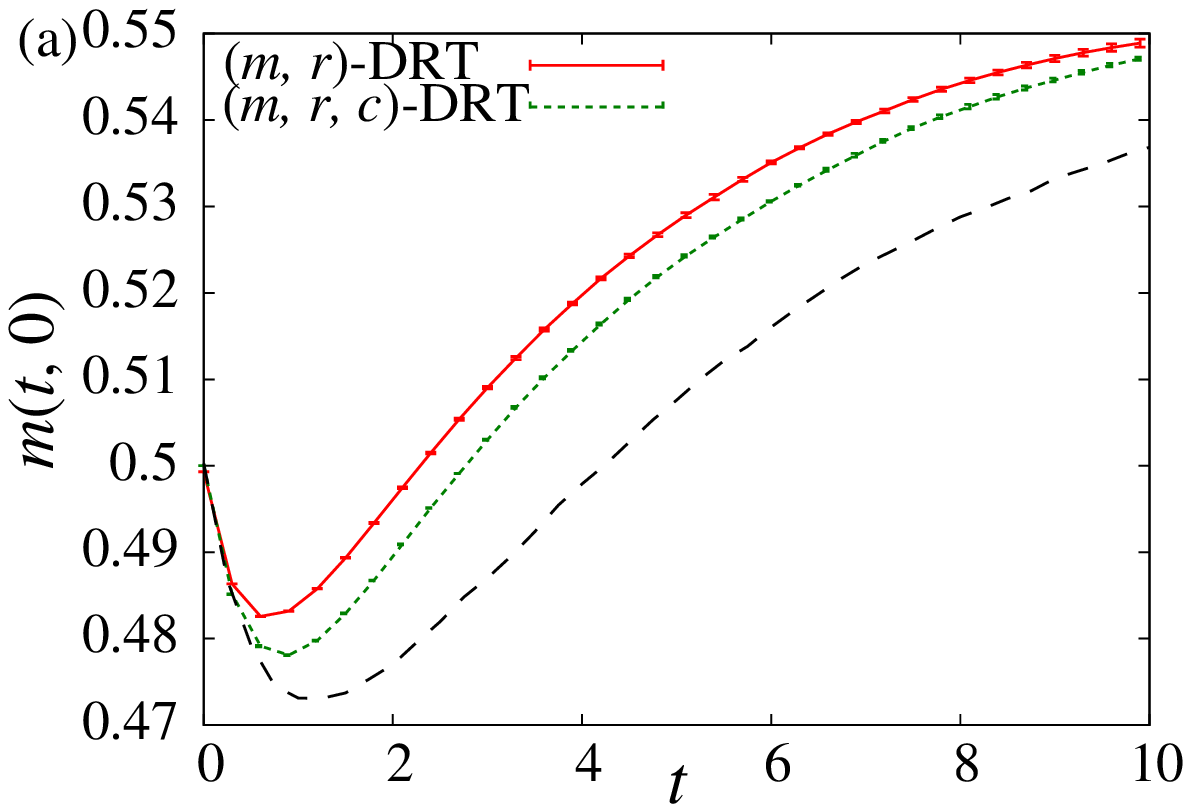}
\includegraphics[width=\figwidth]{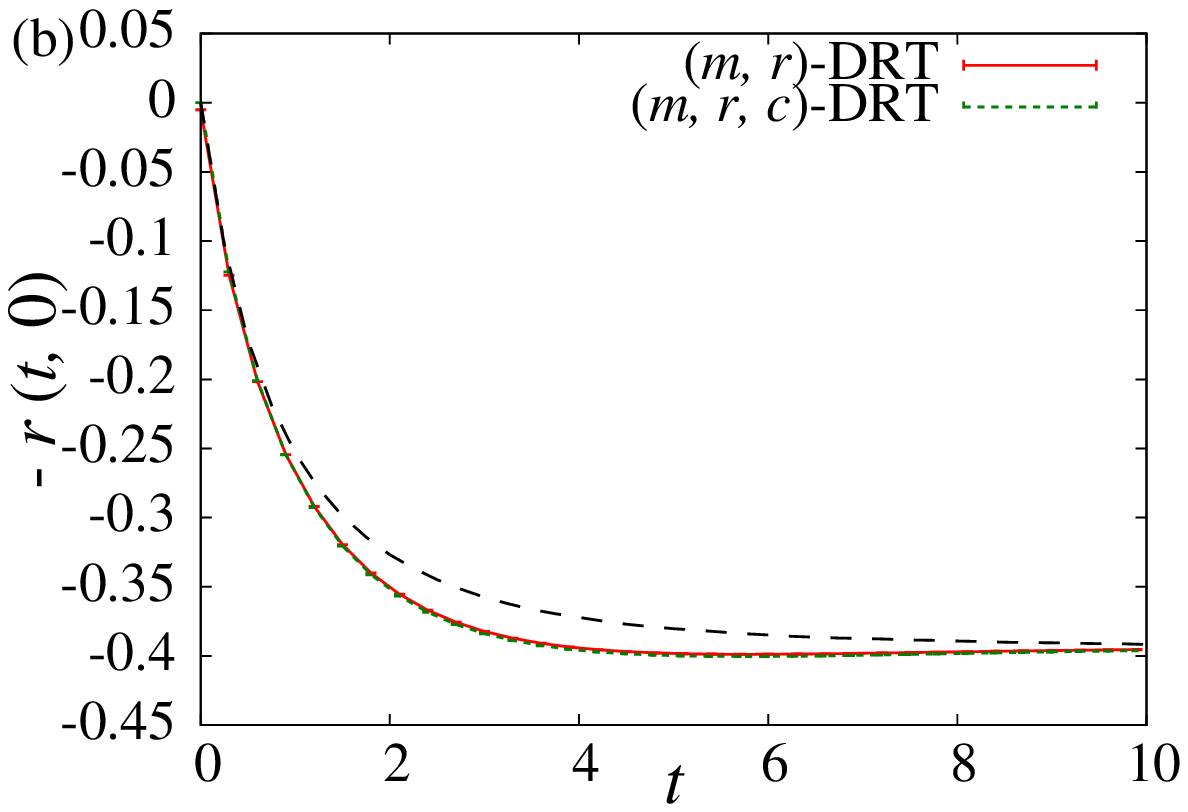}
\end{center}
\caption{(colour online) Time evolution of (a) $m$ and (b) $-r$ at
$T\slash J=1.0$ and $J_0\slash J=1.5$, whose equilibrium state
corresponds to a ferromagnetic phase. The lines are the same as
those in \Fref{fig:para-mr}.
}
\label{fig:ferro-mr}
\begin{center}
\includegraphics[width=\figwidth]{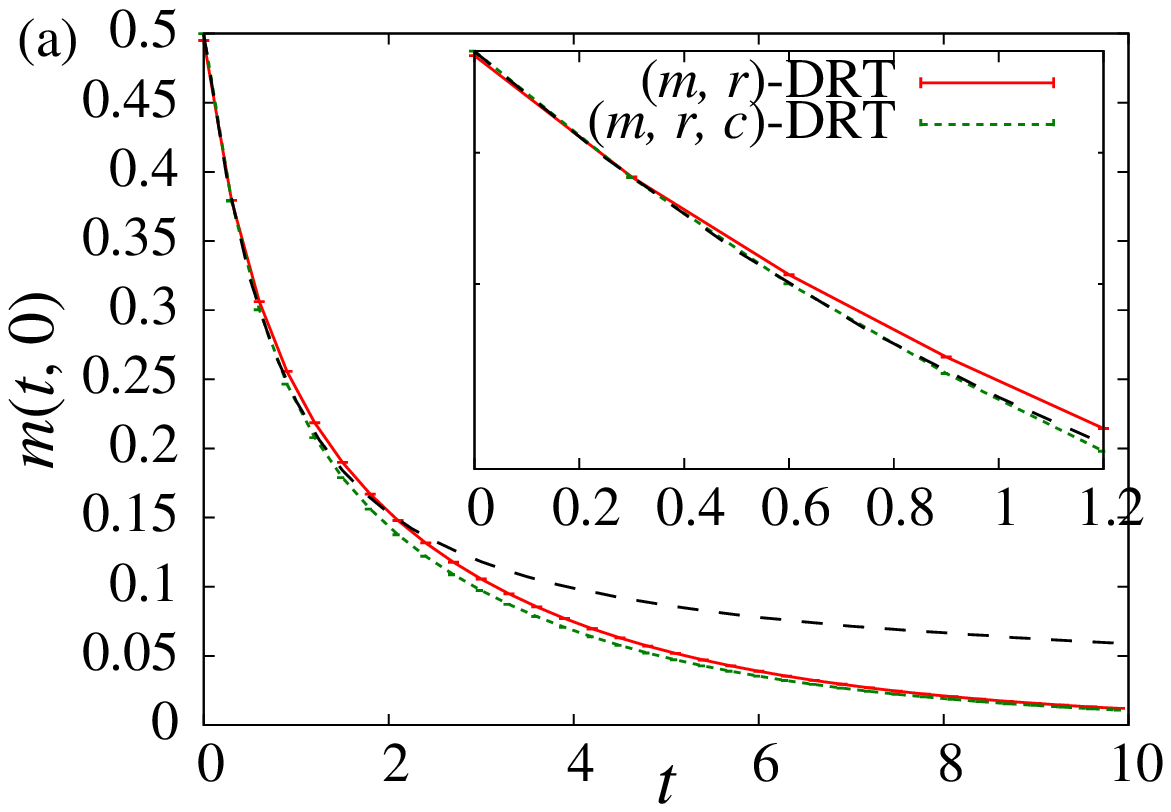}
\includegraphics[width=\figwidth]{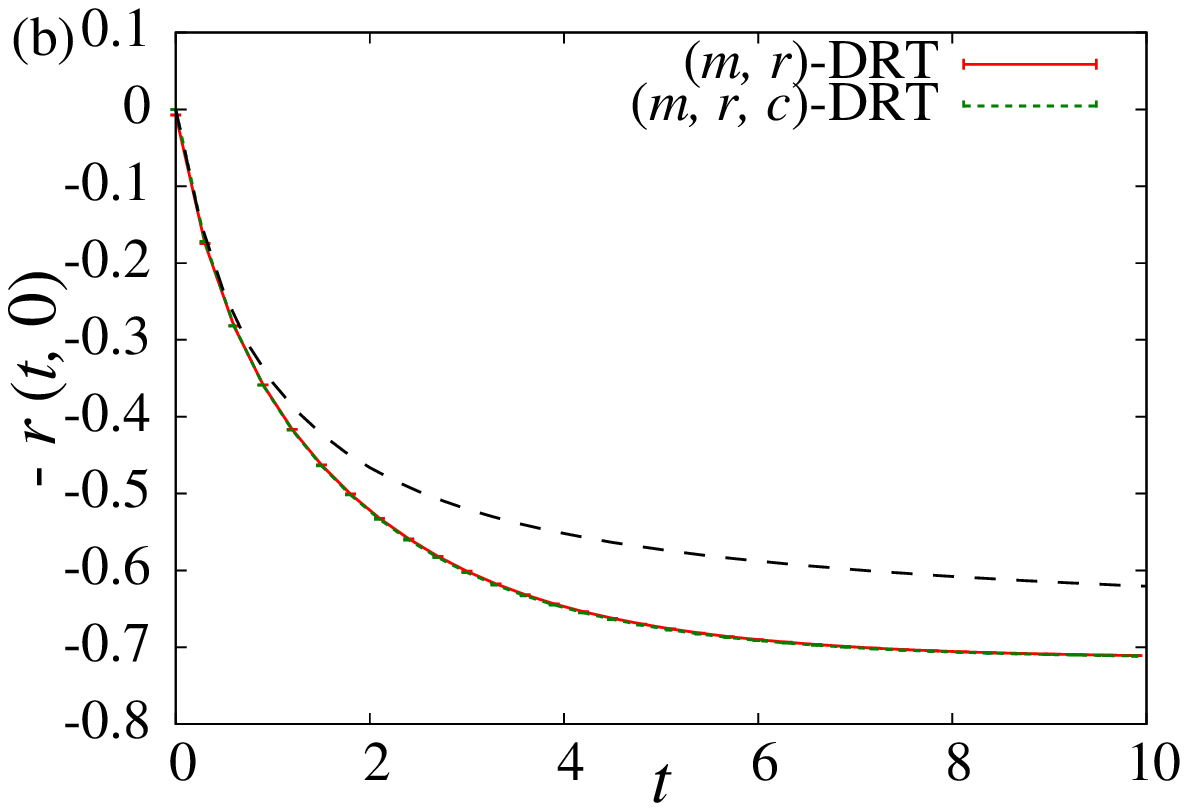}
\end{center}
\caption{(colour online) Time evolution of (a) $m$ and (b) $-r$ at
$T\slash J=0.5$ and $J_0\slash J=0.0$, whose equilibrium state
corresponds to a spin-glass phase. The lines are the same as those
in \Fref{fig:para-mr}. The short time behaviour of the magnetization
is shown in the inset of (a).
}
\label{fig:sg-mr}
\end{figure}

The evolutional dynamics of $m,~r$, and $c$, given by
(\ref{eq:m})--(\ref{eq:c}), are solved numerically. The continuous
evolutional dynamics are treated discretely in the numerical
simulation with a unit step of $\Delta t = 0.01$.
The changes over time, 
$\Delta m =m(t+\Delta t)-m(t)$, $\Delta r =r(t+\Delta t)-r(t)$, and
$\Delta c =c(t+\Delta t)-c(t)$, are obtained by calculating the
noise distribution $D(z,S|\Omega)$ at each time step $t$. To do so,
the equations of the variables $(\mu,\mu_0,\gamma,q,q_0,q_t)$,
(\ref{eq:m_RS})--(\ref{eq:c_RS}) and
(\ref{eq:q_RS})--(\ref{eq:qt_RS}), should be solved under the given
values of the macroscopic quantities
$\Omega(t)=\{m_0,r_0,m(t),r(t),c(t)\}$. The conjugates
$(\mu,\mu_0,\gamma)$ and overlaps $(q,q_0,q_t)$ are solved
alternately until (\ref{eq:m_RS})--(\ref{eq:c_RS}) and
(\ref{eq:q_RS})--(\ref{eq:qt_RS}) are satisfied simultaneously; the
conjugates are implicitly solved by the Newton method based on
(\ref{eq:m_RS})--(\ref{eq:c_RS}) under fixed values of the overlaps,
and the overlaps are solved recursively based on
(\ref{eq:q_RS})--(\ref{eq:qt_RS}) under fixed conjugates. The time
dependence of the conjugates and overlap parameters are shown in
\ref{app:overlap-conjugate}, which correspond to the time evolution
of $m$, $r$ and $c$ shown in this section.

The results of the dynamical equations are compared with a Monte
Carlo (MC) simulation for $N=4096$. The MC simulation settings are
as follows. The initial spin configuration is randomly generated
according to the distribution,
\begin{eqnarray}
P_S(S_i)=\frac{1+m_0}{2}\delta_{S_i,1}+\frac{1-m_0}{2}\delta_{S_i,-1},
\label{eq:P_S}
\end{eqnarray}
and the interaction $\bm{J}$ is generated according to the
distribution \eref{eq:P_J}. The MC dynamics are averaged over 1000
realizations of $\bm{J}$ and 100 initial configurations of $\bm{S}$
for each $\bm{J}$.
A set of randomly generated spin configurations 
and interactions generally provides $r(\bm{S})\sim O(N^{-1})$, which
is sufficiently small to be regarded as zero at $N=4096$. Therefore,
the average over the initial configuration in the MC simulation
corresponds to uniform sampling from the subshell characterized by
$m_0$ and $r_0=0$, and the overlap $q_0$ at time $t_w=0$ is given by
$m_0^2$. In the MC simulation, a randomly chosen $k$-th spin is
flipped with probability $w_k$ given by \eref{eq:transition} every
$1\slash N$ step to match the time scale to that of the DRT.

To clarify the effects of involving the dynamical equation of $c$
into the DRT, the dynamics of $m$ and $r$ are also compared with
that of the original DRT \cite{Coolen-Sherrington}, in which the
subshell is characterized by $m$ and $r$ only. The original DRT and
that including the autocorrelation function are denoted by
$(m,r)$--DRT and $(m,r,c)$--DRT, respectively. The results in three
parameter regions corresponding to paramagnetic, ferromagnetic, and
spin-glass phases are shown here.

\subsection{Time evolution of the magnetization and randomness energy}

The time evolution of $m$ and $-r$ are shown in Figures
\ref{fig:para-mr}, \ref{fig:ferro-mr}, and \ref{fig:sg-mr} for
$J_0\slash J=1.0$ and $T\slash J=2.0$ (paramagnetic phase),
$J_0\slash J=1.5$ and $T\slash J=1.0$ (ferromagnetic phase), and
$J_0\slash J=0.0$ and $T\slash J=0.5$ (spin-glass phase),
respectively. The initial condition at $t_w=0$ is set to $m_0=0.5$
and $r_0=0$.

In the paramagnetic phase (\Fref{fig:para-mr}), the time evolution
of $m$ and $r$ in the MC simulation is well described by the
$(m,r,c)$-DRT proposed here as well as the original $(m,r)$-DRT. The
time evolution does not change significantly by including the
autocorrelation function into the DRT formula in the paramagnetic
phase.

In the ferromagnetic phase (\Fref{fig:ferro-mr}), the time evolution
of $m$ predicted by the $(m,r,c)$-DRT is improved from that of the
original $(m,r)$-DRT. A discrepancy between the MC simulation and
the DRT results appears in early time steps, but it is confirmed
that the fixed point of the DRT is coincident with the MC results as
$t\to\infty$. The time evolution of $r$ described by the
$(m,r,c)$-DRT is almost coincident with that described by the
original $(m,r)$-DRT. At sufficiently large $t$, the DRT predicts
the correct value of $r$, but the decay to the fixed point at around
$2\simleq t\simleq 8$ is faster than the MC result.

\Fref{fig:sg-mr} shows the time evolution of $m$ and $r$ in the
spin-glass phase. The short time behaviour of $m$ in the
$(m,r,c)$-DRT is slightly improved from that in the $(m,r)$-DRT, as
shown in the inset of \Fref{fig:sg-mr} (a). The fixed point of $m$
in both DRTs 
is the same, i.e. the RS solution of the SK
model. The behaviour of $r$ in both DRTs 
is similar, but the long time behaviour in the MC simulation is not
correctly predicted by either DRT.

To summarize, by taking into account the time evolution of the
autocorrelation function, the short time behaviour of the
magnetization is improved in the ferromagnetic and spin-glass
phases, but there is little effect on the randomness energy in any
phase.

\begin{figure}
\begin{center}
\includegraphics[width=\figwidth]{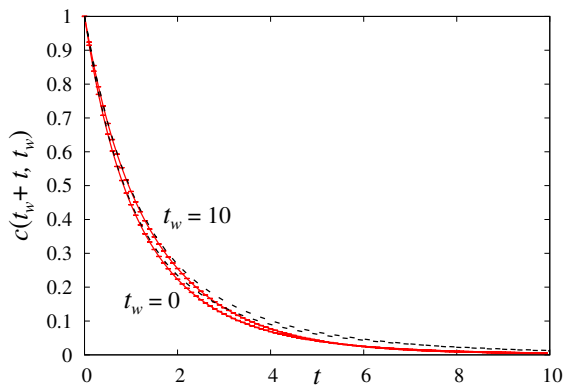}
\end{center}
\caption{(colour online) Time evolution of the autocorrelation
function $c(t_w+t,t_w)$ at $J_0\slash J=1.0$ and $T\slash J=2.0$.
The dashed lines are the results of the MC simulation.}
\label{fig:para-c}
\end{figure}

\subsection{Time evolution of the autocorrelation function}

The time evolution of the autocorrelation function in the
$(m,r,c)$-DRT at a waiting time $t_w$ is compared to the MC
simulation. The autocorrelation $c(t_w+t,t_w)$ for $t_w>0$ is
calculated according to \eref{eq:dc/dt} after resetting
$m_0=m(t_w),~r_0=r(t_w)$, and $q_0=q(t_w)$ where $m(t_w)$ and
$r(t_w)$ are the magnetization and randomness energy at time $t_w$
obtained according to the dynamics \eref{eq:dm/dt} and
\eref{eq:dr/dt} with certain initial conditions $m_0$ and $r_0$. The
overlap at time $t_w$, $q(t_w)$, is determined by \eref{eq:q_RS} for
given values of $m(t_w)$ and $r(t_w)$. Therefore, the
non-equilibrium state at time $t_w$ is assumed to be specified by
three macroscopic quantities: $m(t_w)$, $r(t_w)$, and $q(t_w)$. The
time invariance of the autocorrelation function at sufficiently
large $t_w$ is confirmed over the whole parameter region as
$c(t)=\lim_{t_w\to\infty}c(t_w+t,t_w)$. When one sets $r_0=0$, a
solution of the fixed point equations is given by $\rho_0=\gamma=0$,
as discussed in section \ref{sec:fixed-point}, and in this case the
fixed point of the autocorrelation corresponds to $c^*=q_t=m_0\times
m^*$. Therefore, as $t\to\infty$, if this fixed-point solution is
stable, the autocorrelation decreases to zero in the paramagnetic
and spin-glass phases irrespective of the value of $m_0$. On the
other hand, it converges to a finite value given by $m_0m^*$ in the
ferromagnetic phase.

\Fref{fig:para-c} shows the time evolution of $c(t_w+t,t_w)$ at
$J_0\slash J=1.0$ and $T\slash J=2.0$, in which the equilibrium
state corresponds to the paramagnetic phase. Both of the results for
$t_w=0$ and $t_w=10$ show good agreement with the results of the MC
simulation. The autocorrelation decreases to zero as time increases
for any $t_w$. In the paramagnetic phase, the solution $c^*=q_t=0$
is a stable solution of the fixed point equations.

\begin{figure}
\begin{center}
\includegraphics[width=\figwidth]{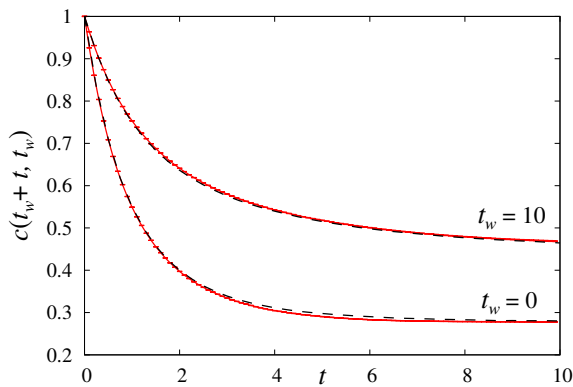}
\end{center}
\caption{(colour online) Time evolution of the autocorrelation
function $c(t_w+t,t_w)$ at $J_0\slash J=1.5$ and $T\slash J=1.0$.
The dashed lines are the results of the MC simulation.}
\label{fig:ferro-c}
\end{figure}

In \Fref{fig:ferro-c}, the time evolution of $c(t_w+t,t_w)$ for
$t_w=0$ and $t_w=10$ are shown for the ferromagnetic phase
($J_0\slash J=1.5$ and $T\slash J=1.0$). The dynamics of the
autocorrelation function are well described by the $(m,r,c)$-DRT for
both $t_w$ values. As the waiting time $t_w$ increases, the
relaxation of the autocorrelation appears to be slow.
The converged value
of $c(t_w+t,t_w)$ corresponds to $m(t_w)\times m(t_w+t)$ for
both $t_w$ times. In the ferromagnetic phase, $c^*=q_t=m_0m^*$ is a
stable solution of the fixed point equations where $m^*\sim0.56$ in
this parameter region.

\begin{figure}
\begin{center}
\includegraphics[width=\figwidth]{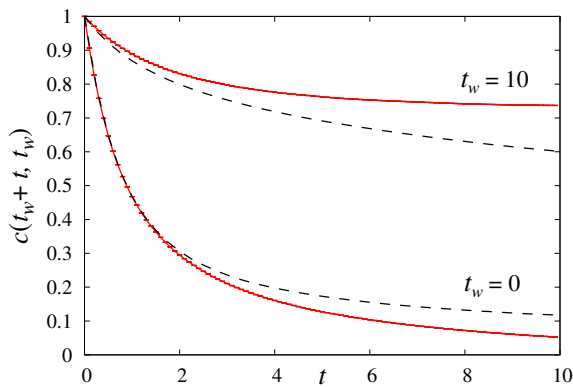}
\end{center}
\caption{(colour online) Time evolution of the autocorrelation
function $c(t_w+t,t_w)$ at $J_0\slash J=0$ and $T\slash J = 0.5$.
The dashed lines are the results of the MC simulation.
}
\label{fig:sg-c}
\end{figure}

In \Fref{fig:sg-c}, the time evolution of the autocorrelation
function in the spin-glass phase, $J_0\slash J=0$ and $T\slash
J=0.5$, for $t_w=0$ and $t_w=10$ are shown. At $t_w=0$, the short
time behaviour of the autocorrelation $c$ is well described by the
$(m,r,c)$-DRT, but it deviates from the results of the MC simulation
as $t$ increases. The fixed point of the autocorrelation function
for $t_w=0$ in the spin-glass phase is $c^*=0$, as in the
paramagnetic phase. As $t_w$ increases, the discrepancy between the
autocorrelation function in the $(m,r,c)$-DRT and that in the MC
simulation widens even in short time steps. In the spin-glass phase,
the initial conditions at $t_w$, $m_0=m(t_w)$, $r_0=r(t_w)$, and
$q_0=q(t_w)$ predicted by the $(m,r,c)$-DRT, are not coincident with
the actual values as $t_w$ increases, and hence the dynamics after
$t_w$ in the $(m,r,c)$-DRT do not agree with those in the MC
simulation. Generally, DRT predicts a faster relaxation of $m$ and
$r$ than the actual relaxation as shown in \Fref{fig:sg-mr}, which
indicates that DRT describes the dynamics at time steps later than
the actual time. Therefore, the relaxation of the dynamical equation
in DRT at $t_w>0$ is slower than that in the MC simulation. By
setting the initial conditions of $m(t_w)$ and $r(t_w)$ in the
$(m,r,c)$-DRT to those of the actual values in the MC simulation at
$t_w$, the short time behaviour of $c(t_w+t,t_w)$ is improved,
although there is still a discrepancy between the results.
By comparing the
results in the paramagnetic and spin-glass phases, it is found that
the $(m,r,c)$-DRT can describe the slower relaxation of the
autocorrelation function in the spin-glass phase than that in
paramagnetic phase.

\begin{figure}
\begin{center}
\includegraphics[width=\figwidth]{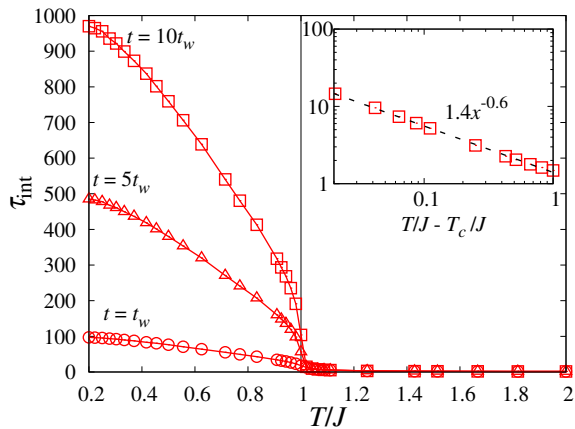}
\end{center}
\caption{(colour online)
The temperature dependence of the integral relaxation time $\tau_{\rm int}$
at $J_0\slash J=0$ and $h=0$ for $t_w=100$.
The vertical line at $T_c\slash J=1.0$
indicates the spin-glass transition temperature.
The inset shows the dependence of $\tau_{\rm int}$
on $(T-T_c)\slash J$ with a fitting function denoted by the dashed line.
}
\label{fig:relaxation}
\end{figure}

To characterize the relaxation of the autocorrelation function,
it is convenient to introduce the integral relaxation time defined by
\begin{eqnarray}
\tau_{\rm int}(t_w)=\int_0^t dt' C(t_w+t',t_w).
\end{eqnarray}
The tendency of $\tau_{\rm int}(t_w)$ as $t\to\infty$ is numerically
checked by changing the integration range $t$. \Fref{fig:relaxation}
shows the temperature dependence of $\tau_{\rm int}(t_w)$
at
$J_0\slash J=0$ and $h=0$ for $t_w=100$
with integration ranges $t=t_w$, $t = 5t_w$, and $t = 10t_w$.
The spin-glass transition temperature in
this parameter region is $T_c\slash J=1$, as indicated by the vertical
line in \Fref{fig:relaxation}.
The integral relaxation time in the paramagnetic phase ($T\slash
J>1$) does not depend on the integration range $t$, but it increases
in the spin-glass phase ($T\slash J<1$) as $t$ increases. This means
that $\tau_{\rm int}$ diverges as $t\to\infty$ in the spin-glass
phase. Around the spin-glass transition temperature, $\tau_{\rm
int}$ exhibits a power law behaviour with respect to $T\slash
J-T_c\slash J$ as shown in the inset of \Fref{fig:relaxation}. In
the DRT formulation, $\tau_{\rm int}$ goes to infinity at $T_c$ for
$\tau_{\rm int}(t_w)=1.4(T\slash J-T_c\slash J)^{-0.6}$ at
sufficiently large $t_w$ and $t$, which indicates that the
$(m,r,c)$-DRT can detect the critical slowing down. However, the
exponent of 0.6 is smaller than the expected value of 2.0
\cite{Kirkpatrick-Sherrington}. The proposed $(m,r,c)$-DRT
qualitatively describes slow relaxation in the spin-glass phase and
the critical behaviour around the spin-glass transition temperature,
but further development is necessary for a more accurate
description.


\section{Summary and discussion}
\label{Summary}

In this paper, I have examined the relaxation dynamics 
in the SK model using DRT and including the autocorrelation
function. The joint probability of the microscopic states at times
$t_w$ and $t_w+t$ was introduced, whose time evolution is described
by Glauber dynamics. The microscopic states at both times are
characterized by macroscopic quantities; the magnetization and
energy contributed by randomness at time $t_w+t$ denoted by $m$ and
$r$, those at time $t_w$ denoted by $m_0$ and $r_0$, as well as the
autocorrelation function $c$. Following the DRT procedures, closed
dynamical equations were derived based on the self-averaging and
subshell equipartitioning assumptions. The dynamical equations are
governed by three overlap parameters $q$, $q_0$, and $q_t$ and the
conjugates $\mu,\mu_0$, and $\gamma$ under the RS assumption.
%

The time evolution of $m$ in the $(m,r,c)$-DRT is improved in the
ferromagnetic and spin-glass phases compared with $(m,r)$-DRT
but that 
of $r$ does not change significantly
in any phase. In the paramagnetic and
ferromagnetic phases, the time evolution of the autocorrelation
function $c(t_w+t,t_w)$ was well described by the proposed framework
even at a finite $t_w$. In the spin-glass phase, the short time
behaviour of $c(t_w+t,t_w)$ at $t_w=0$ was well described by the
$(m,r,c)$-DRT, but the theoretical prediction deviated from the MC
simulation as the waiting time $t_w$ and the difference between the
two times $t$ increased.
The temperature dependence of $c(t_w+t,t_w)$ was characterized by
the integral relaxation time $\tau_{\rm int}$, and it was found that
slow relaxation in the spin-glass phase and the critical slowing down are
qualitatively described by the current $(m,r,c)$-DRT formulation.


To improve the results in the spin-glass phase, the validity of the
assumptions, RS and subshell equipartitioning with some macroscopic
quantities, should be discussed. A promising way for the improvement
is to introduce a higher-order macroscopic quantity to characterize
the subshell as in \cite{Laughton} and RSB \cite{beyond,Nishimori}.
Another straightforward development is to take into account
multitime correlations. One can obtain the dynamical equations of
the multitime correlation function by starting with a joint
probability distribution of the microscopic states at several time
steps. In this case, the analytical method that is employed to
derive the dynamical equations corresponds to the replica method for
a multi-replicated system consisting of spin systems at each time
step, which is called real-replica analysis and employed to
understand the RSB picture in the equilibrium state
\cite{Kurchan-Parisi-Virasoro}.

DRT has been developed here to include the dynamical equation of the
autocorrelation function defined by the microscopic states at two
specific times. From this study, the ability and prospects of DRT
for describing the dynamics of random systems has been shown. A
modification of the DRT proposed in this paper will provide an
analytical description of the experimentally and numerically
employed protocols for glassy systems. For example, it is expected
that the fluctuation-dissipation relation can be checked in the
present formulation by comparing the time evolutions of the
susceptibility and autocorrelation function \cite{Crisanti-Ritort}.
The current theory is applicable to a $p$-body interaction system;
in this case the relationship between bifurcation in dynamical
equations and static phase transition associated with RSB should be
discussed \cite{Sakata}.

\ack

I would like to thank A.C.C. Coolen, K. Hukushima, Y. Kabashima, and
T. Nakajima for helpful comments and discussions. This work was
partially supported by a Grant-in-Aid for JSPS Fellows (No.
23--4665).

\appendix
\section{Overlap and conjugate parameters}
\label{app:overlap-conjugate}

\begin{figure}
\begin{center}
\includegraphics[width=2.in]{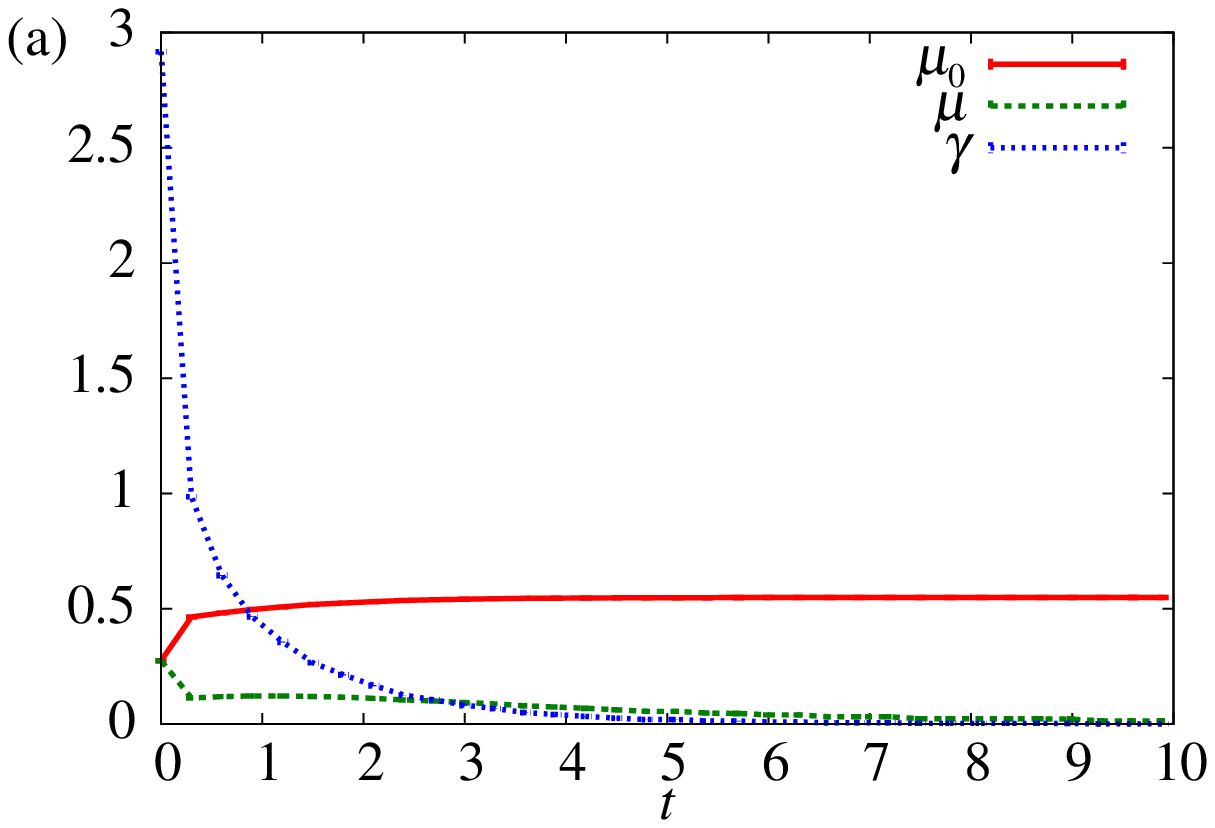}
\includegraphics[width=2.in]{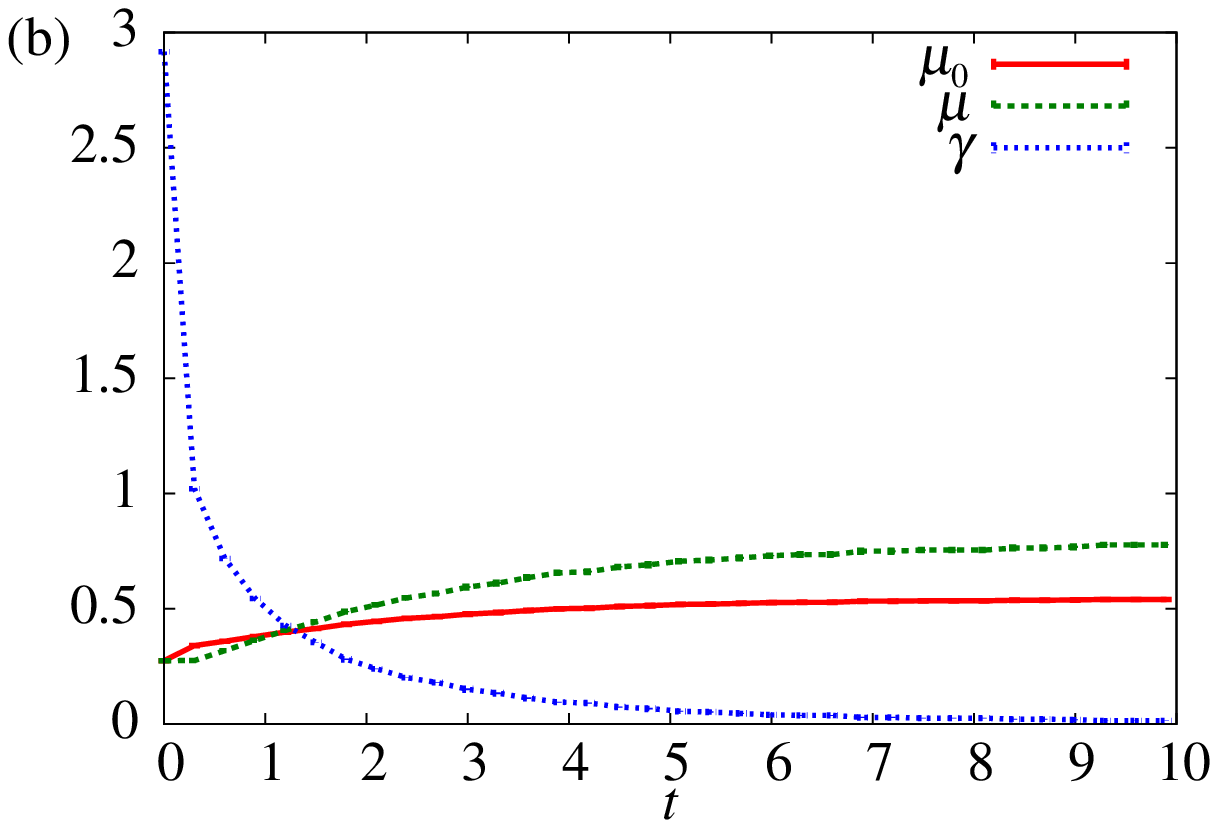}
\includegraphics[width=2.in]{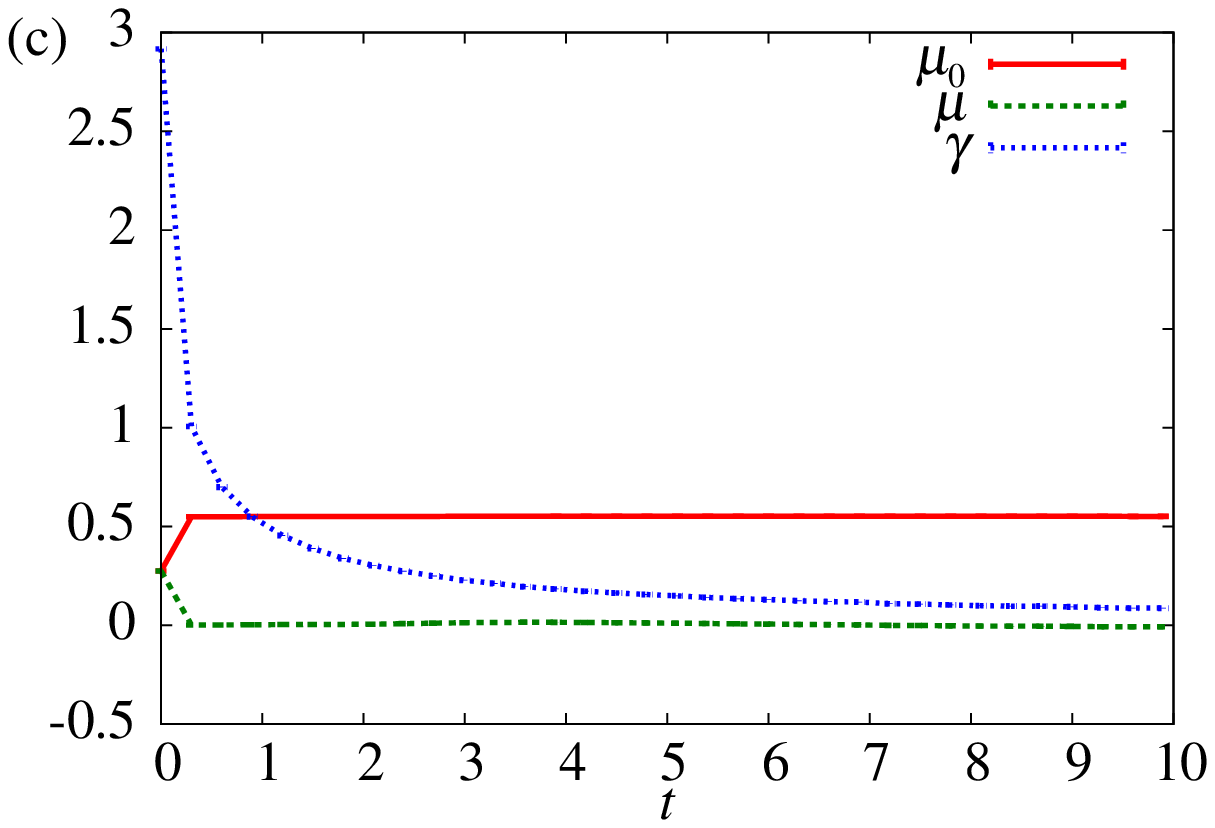}
\end{center}
\caption{(colour online) Time dependence of the conjugate parameters
at (a) $T\slash J=2.0$ and $J_0\slash J=1.0$, (b) $T\slash J = 1.0$
and $J_0\slash J = 1.5$, and (c) $T\slash J = 0.5$ and $J_0\slash J
= 0.0$ for $t_w=0$.}
\label{fig:conjugate}
\begin{center}
\includegraphics[width=2.in]{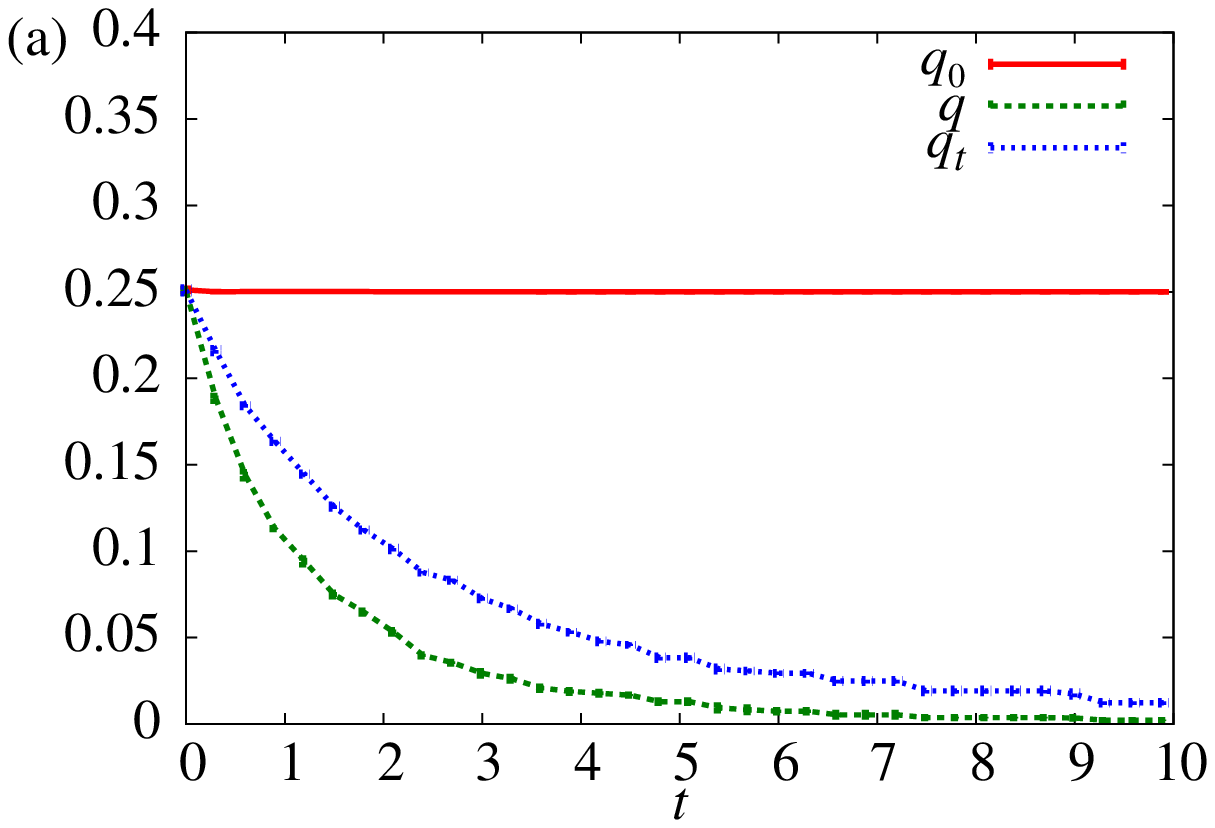}
\includegraphics[width=2.in]{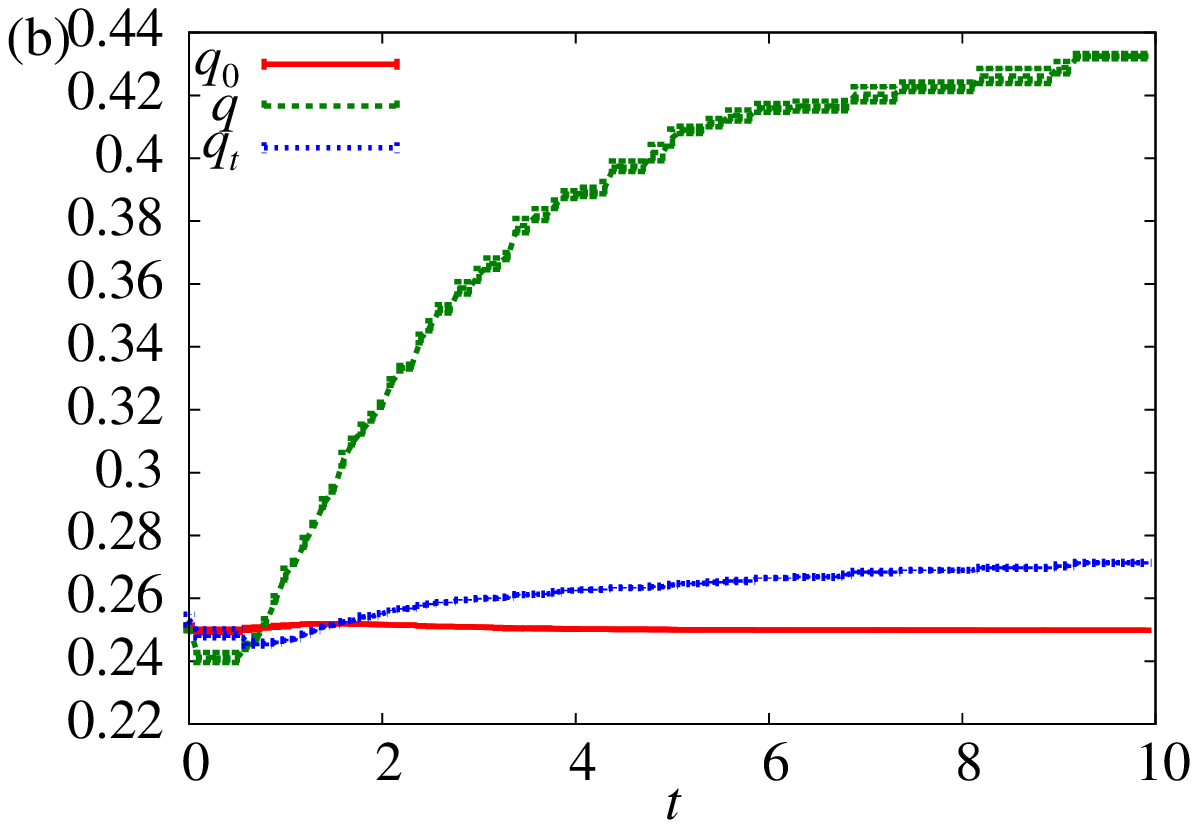}
\includegraphics[width=2.in]{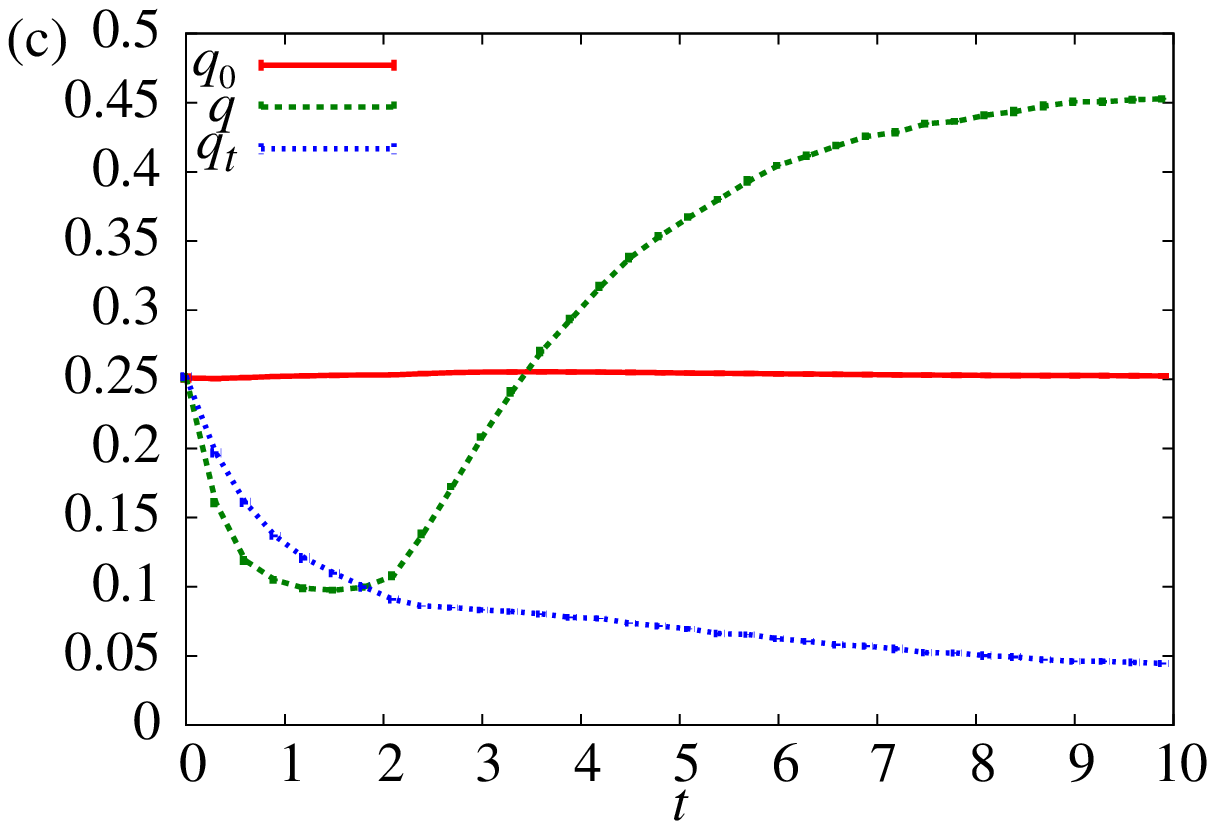}
\end{center}
\caption{(colour online) Time dependence of the overlap parameters
at (a) $T\slash J=2.0$ and $J_0\slash J=1.0$, (b) $T\slash J = 1.0$
and $J_0\slash J = 1.5$, and (c) $T\slash J = 0.5$ and $J_0\slash J
= 0.0$ for $t_w=0$.}
\label{fig:qs}
\end{figure}

To obtain the time evolution of the macroscopic variables, the
overlaps $q_0,q,q_t$ and conjugates $\mu_0,\mu,\gamma$ should be
solved at each time step. \Fref{fig:conjugate} and \ref{fig:qs} show
the time dependence of the conjugates and overlap parameters,
respectively, at $t_w=0$. The initial condition is given by
$m_0=0.5$ and $r_0=0$, and the corresponding time evolution of the
macroscopic quantities are shown in section \ref{results}.
As shown in \Fref{fig:conjugate}, the conjugate $\gamma$ decreases
to 0 with increasing time. As $t\to\infty$, the conjugates $\mu_0$
and $\mu$ converges to the values given by $\tanh^{-1}m_0$ and
$\beta (J_0m^*+h)$, respectively. The causality that $q_0$ takes a
constant value (in this case given by $m_0^2$) is achieved at all
time steps by appropriately controlling the conjugates as shown in
\Fref{fig:qs}. The overlap between times $t_w$ and $t_w+t$, $q_t$,
decreases to zero in the paramagnetic and spin-glass phases and
converges to $m_0\times m(t)$ in the ferromagnetic phase.

\end{document}